\documentclass[aps,prb,reprint,groupedaddress,showpacs,floatfix]{revtex4-1}

\usepackage[utf8]{inputenc}
\usepackage{color}
\usepackage{graphicx}

\usepackage{amsmath}
\usepackage{graphicx}
\graphicspath{{./images/}}

\begin{document}

\title{Quantum model of coupled intersubband plasmons}

\author{G. Pegolotti, A. Vasanelli}
\email[]{angela.vasanelli@univ-paris-diderot.fr}
\author{Y. Todorov and C. Sirtori}
\affiliation{Universit\'e Paris Diderot, Sorbonne Paris Cit\'e, Laboratoire Mat\'eriaux et Ph\'enom\`enes Quantiques, UMR7162, 75013 Paris, France}

\date{\today}

\begin{abstract}
We present a quantum model to calculate the dipole-dipole coupling between electronic excitations in the conduction band of semiconductor quantum wells. We demonstrate that the coupling depends on a characteristic length, related to the overlap between microscopic current densities associated with each electronic excitation. As a result of the coupling, a macroscopic polarization is established in the quantum wells, corresponding to one or few bright collective modes of the electron gas. Our model is applied to investigate the interplay between tunnel coupling and Coulomb interaction in the absorption spectrum of a dense electron gas. 
\end{abstract}


\maketitle
\section{Introduction}
Intersubband transitions in quantum wells are widely exploited in optoelectronic devices for the mid- and far-infrared domain, such as quantum cascade lasers~\cite{faist} and detectors.~\cite{gendron} The properties of these devices can be described in a single particle picture~\cite{helm} and are essentially based on two quantum phenomena: electronic confinement and tunnelling. However, intersubband transitions are an intrinsically collective phenomenon, involving large densities of interacting electrons.~\cite{ando} The most important manifestation of this collective character is the fact that, in the presence of electromagnetic radiation, each electron feels an effective field induced by the excitation of the other electrons, called depolarization field.~\cite{helm, ando} When a single subband is occupied, this results in a blue-shift of the absorption peak with respect to the transition frequency, corresponding to the excitation of a collective mode of the system, the intersubband plasmon.~\cite{wendler} The effect of the depolarization field is even more spectacular in a highly doped quantum well, when several subbands are occupied. In this case, the measured optical spectrum consists of a single resonance, whose energy is completely different with respect to the bare intersubband transitions.~\cite{delteil_PRL2012} This resonance corresponds indeed to the excitation of a collective mode of the system, the \emph{multisubband plasmon}, resulting from the phase locking of all the different intersubband transitions.   

Even though the study of many-body effects in the optical response of an electron gas has been the object of a vaste literature,~\cite{ando,allen,nikonov,pinczuk,vinter,warburton_PRL,wendler,zaluzny,li} its application to optoelectronic intersubband devices is very limited.~\cite{liu_THzqwip} Yet, devices fully exploiting the collective character of intersubband transitions have an enourmous potential, as intersubband and multisubband plasmons have the properties of a superradiant state, issued from the coherent superposition of a huge number of oscillating dipoles.~\cite{gross, sdl_PRB2012} In order to develop new devices based on these states, a new quantum engineering of the collective, rather than  the single particle, states has to be constructed. In particular, one has to investigate the interplay between tunnel coupling, one of the fundamental features of intersubband devices, and the coherence induced by Coulomb interaction.

The aim of this work is to provide a quantum model describing the coupling between intersubband plasmons in a single quantum well and in systems of tunnel-coupled wells and its consequences on the infrared absorption spectra. Our approach, suitable when the depolarization field is the dominant many-body contribution, relies on the use of the dipole representation~\cite{todorov_PRB1, todorov_PRB2} to treat the light-matter interaction, which naturally accounts for the depolarization field. 

We first present our model in Sec.~\ref{sec:theory} and define the relevant quantities for the calculation of the absorption spectra. In particular, we define for each intersubband transition a microscopic current density, which can be related to the oscillator strength of the transition. We demonstrate that the overlap between these currents leads to the definition of a typical length for the Coulomb interaction, hence determining the coupling between intersubband plasmons. The diagonalization of the electronic Hamiltonian allows us to define a microscopic current for each collective mode and to calculate the absorption spectrum. A sum rule is established, expressing the conservation of the total interaction with light when going from the single particle to the many-body picture. In Sec.~\ref{sec:single_well}, the theory is applied to the case of a single quantum well with several occupied subbands. We demonstrate that the semiclassical model describing the intersubband absorption in terms of Drude-Lorentz oscillators is recovered as a special case of our formalism, when all the overlap integrals between intersubband currents are comparable. Then, in Sec.~\ref{sec:at}, the absorption spectrum of a dense electron gas in a system of asymmetric wells is investigated. In this case, we demonstrate that the coupling between a bright and an almost dark transition results in an optical spectrum composed of two peaks of equal amplitude. Here the coherence induced by Coulomb interaction gives rise to a phenomenon similar to the Autler Townes effect.~\cite{at_pr100_1955, cohen_book} Finally, we explore in Sec.~\ref{sec:coupledMPS} a system of two tunnel-coupled quantum wells with several occupied subbands. The resulting absorption spectrum is interpreted in terms of the collective currents of the multisubband plasmons of the system.

\section{Theoretical model}
\label{sec:theory}
This section is devoted to the presentation of our quantum model, describing the optical properties of a dense electron gas when the depolarization field is the dominant many-body contribution. Our model is based on the formalism developed in Ref.~\onlinecite{todorov_PRB1}, which takes advantage of the fact that a treatment of the light-matter coupling in the dipole representation naturally accounts for the depolarization effect, arising from dipole-dipole Coulomb interaction. Indeed, in this representation the interaction Hamiltonian is in general written as, neglecting the magnetic interactions~\cite{cohen_patomes, scheel_book}:
\begin{equation}
\label{int_hamiltonian}
H_{\rm int}=\int{\frac{1}{\varepsilon_0 \varepsilon_s} \left( -\mathbf{D} \cdot \mathbf{P} +\frac{1}{2} \mathbf{P}^2\right) {\rm{d}^3}\mathbf{r}}=H_{l-p}+H_{p}
\end{equation}
where $\varepsilon_s$ is the background dielectric constant, $\mathbf{P}$ is the polarization density operator and $\mathbf{D}$ is the displacement field operator. The Hamiltonian~(\ref{int_hamiltonian}) is composed of two terms: $H_{l-p}$ describes the interaction between the matter polarization and the electromagnetic field, while $H_p$ describes the polarization self-interaction. 

In order to use this representation to calculate the optical properties of the electron gas, it is necessary to define an intersubband polarization density, associated with the excitation of an ensemble of intersubband oscillators along the growth direction. The dipole-dipole interaction between intersubband oscillators, hence the depolarization field, is described by the $H_p$ term. 

Following this approach, we will first define the intersubband polarization and write it in terms of microscopic current densities describing the different intersubband transitions. We will then introduce the intersubband plasmons, issued from the dipole-dipole coupling between intersubband oscillators of the same transition. We will then consider the mutual interaction between intersubband plasmons, resulting in new collective modes of the system, the multisubband plasmons. Finally, we will establish a link between the microscopic current density and the absorption spectrum.

\subsection{Intersubband transitions and single particle absorption spectrum}\label{sec:ISB_transition}
Let us consider a quantum well or a system of tunnel-coupled quantum wells, grown along the direction $z$. We call $\psi_i(z)$ the electronic wavefunctions and $\omega_i$ the corresponding eigenfrequencies. To simplify the notation, we consider a constant effective mass $m^*$ throughout the structure. The generalization to a piecewise constant effective mass is straightforward: all the numerical calculations in this work have been performed with a different effective mass for barriers and wells.

For each intersubband transition $\alpha \equiv i\rightarrow j$ (approximated as vertical in the reciprocal space, neglecting the photon wavevector), of frequency $\omega_\alpha=\omega_j-\omega_i$, we define a bosonic operator $b_\alpha$ such that the matter part of the Hamiltonian is written as~\cite{ciuti_PRB2005}:
\begin{equation}
H_{e}=\sum_\alpha{\hbar \omega_\alpha b_\alpha^\dagger b_\alpha}
\end{equation}
For simplicity, we do not write the sum over the in-plane wavevector of the electronic excitations explicitly. 
Following Ref.~\onlinecite{todorov_PRB1}, we also introduce a current operator $\widehat{j}_z$ in the growth direction $z$ (because of the intersubband absorption selection rule), expressed in terms of the intersubband transition operators:
\begin{align} \nonumber
\widehat{j}_z&=i\frac{e \hbar}{2 m^*\sqrt{S}}\sum_\alpha \xi_\alpha(z) \sqrt{\Delta N_\alpha} \Big( b_\alpha-b_\alpha ^\dagger \Big)=\\
&=i \sum_\alpha{j_\alpha(z)\Big( b_\alpha-b_\alpha ^\dagger \Big)}
\end{align}
where $S$ is the area of the system, $\Delta N_\alpha$ is the surface density of the electronic excitations between the two subbands (i.e. $\Delta N_\alpha=N_i-N_j$) and $\xi_\alpha(z)$ is given by
\begin{equation}\label{eq:ucurrent}
\xi_\alpha(z) \equiv \xi_{ij}(z) = \psi_i(z)\frac{\partial\psi_j(z)}{\partial z} - \psi_j(z)\frac{\partial\psi_i(z)}{\partial z}
\end{equation}
The quantity $j_\alpha(z)$ is a current per unit surface associated with each intersubband transition $\alpha$. Its spatial variation is determined by $\xi_\alpha(z)$, which can thus be considered as a current distribution. Analogously, one can introduce a function $\rho_\alpha$ describing the spatial distribution of the charge oscillating at the frequency of the transition $\alpha$: $\rho_\alpha=\frac{\partial\xi_\alpha (z)}{\partial z}$. 

By using the expression of the current operator, it is possible to define an intersubband polarization operator $\widehat{P}_z$, such that the current is given by its time evolution:
\begin{equation}
\frac{\partial \widehat{P}_z}{\partial t}=\widehat{j}_z=\frac{1}{i\hbar}\left[\widehat{P}_z,H_e \right]
\end{equation}
The intersubband polarization contains contributions from all the intersubband transitions and can be written as:
\begin{align}\nonumber
\widehat{P}_z&=\frac{e \hbar}{2 m^*\sqrt{S}}\sum_\alpha {\frac{\xi_\alpha(z)}{\omega_\alpha}\sqrt{\Delta N_\alpha}\Big(b_\alpha^\dagger + b_\alpha \Big)} =\\
&= \sum_\alpha {\frac{j_\alpha(z)}{\omega_\alpha}\Big(b_\alpha^\dagger + b_\alpha \Big)}
\end{align}
Note that the spatial variation of the polarization operator is also described by the current distribution $\xi_\alpha(z)$. Interestingly, this quantity can be easily related to the dipole matrix element $z_\alpha$ of the transition $\alpha$:
\begin{equation} 
\int_{-\infty}^{+\infty}{\xi_\alpha(z)\mathrm{d}z}=\int_{-\infty}^{+\infty}{\rho_\alpha(z)\, z \mathrm{d}z}=\frac{2 m^* \omega_\alpha}{\hbar} z_\alpha.
\end{equation}
This observation is very important for the following part of this work, as it establishes a link between the optical properties of the electron gas and the intersubband current densities. Indeed, the integral of the current density is related to the dipole matrix element through:
\begin{equation}\label{eq:j_integral}
\frac{S}{e^2\omega_\alpha}\left\vert\int_{-\infty}^{+\infty}{j_\alpha(z)\mathrm{d}z}\right\vert^2=  \omega_\alpha\vert z_\alpha\vert^2  \Delta N_\alpha
\end{equation}
As the dipole matrix element also determines the strength of the interaction with light, Eq.~(\ref{eq:j_integral}) allows establishing a relation between the current density integral and the absorption spectrum. 

In a single particle picture, the absorption coefficient is calculated as:~\cite{helm}
\begin{align}\nonumber
\alpha_{2D}(\omega)&=\frac{e^2 \hbar}{2 \epsilon_0 c m^* \sqrt{\epsilon_s}}\sum_\alpha {f_\alpha \Delta N_\alpha \textit{L}(\omega-\omega_\alpha)}=\\
&=\frac{e^2}{\epsilon_0 c \sqrt{\epsilon_s}}\sum_\alpha {\omega_\alpha \left \vert z_\alpha \right \vert^2 \Delta N_\alpha \textit{L}(\omega-\omega_\alpha)}
\end{align}
where $f_\alpha$ is the oscillator strength of the transition, $c$ is the speed of light, and $\textit{L}(\omega-\omega_\alpha)$ is a Lorentzian centered at the intersubband transition frequency. By using Eq.~(\ref{eq:j_integral}), the absorption coefficient can be related to the current densities through:
\begin{eqnarray}\label{eq:abs_spectrum_ISB}
\alpha_{2D}&=&\frac{e^2 S}{\epsilon_0 c \sqrt{\epsilon_s}} A(\omega)\\ \label{eq:abs_spectrum_ISB_2}
A(\omega) &=& \sum_\alpha {\frac{1}{\omega_\alpha } \left \vert \int_{-\infty}^{+\infty}{j_\alpha(z)\mathrm{d}z} \right \vert^2  \textit{L}(\omega-\omega_\alpha)}
\end{eqnarray}
This result is detailed in the Appendix, where the Fermi golden rule is applied to calculate the absorption spectrum in the dipole representation. Physically, it expresses the relation between the a.c. currents associated with the different optically active intersubband transitions and the total absorption of the electron gas.

\subsection{From intersubband transitions to multisubband plasmons}
In order to consider the effect of the depolarization field, we diagonalize the Hamiltonian $H_{\rm{plasmon}}=H_{e}+H_p$, which physically describes the ensemble of interacting intersubband dipolar oscillators.
$H_p$ is expressed in terms of the intersubband excitation operators as:~\cite{todorov_PRB1}
\begin{equation}
H_p=\frac{e^2}{2 \varepsilon_0 \varepsilon_s}\sum_{\alpha,\beta}{S_{\alpha,\beta}\sqrt{\Delta N_\alpha \Delta N_\beta}\Big( b_\alpha^\dagger + b_\alpha \Big)\Big( b_\beta^\dagger + b_\beta \Big)}
\end{equation}
where $S_{\alpha,\beta}$ is given by:
\begin{equation}
 S_{\alpha\beta} = \frac{1}{\hbar\omega_\alpha}\frac{1}{\hbar\omega_\beta} \left(\frac{\hbar^2}{2m^*}\right)^2\int_{-\infty}^{+\infty}\mathrm{d} z\, \xi_\alpha(z) \xi_\beta(z)
\end{equation}
$S_{\alpha\beta}$ defines a characteristic length, depending on the overlap between microcurrents. Diagonal terms $S_{\alpha\alpha}$ refer to the interaction between dipoles associated with the same transition, while off-diagonal terms $S_{\alpha\beta}$ refer to dipoles belonging to different transitions. Note that this expression is equivalent to the Coulomb tensor:~\cite{ando, allen, vinter_PRB1977}
\begin{align}\nonumber
 S_{\alpha\beta} \equiv S_{ij,kl} =& \int_{-\infty}^{+\infty}\mathrm{d} z\, \left[\int_{-\infty}^z \mathrm{d}z'\,\psi_i(z')\psi_j(z')\right]\\\label{eq:def_S}
&\times\left[\int_{-\infty}^z \mathrm{d}z''\,\psi_k(z'')\psi_l(z'')\right]
\end{align}

The diagonalization of $H_{\rm{plasmon}}$ is done in two steps. We first consider the interaction between oscillators corresponding to the same intersubband transition ($\alpha=\beta$), resulting in the usual intersubband plasmons.~\cite{todorov_PRL2010} Secondly, the coupling between these intersubband plasmons is considered, to finally obtain the new eigenmodes of the system, the multisubband plasmons.

The destruction operators of the intersubband plasmons are expressed as:
\begin{equation}
p_\alpha=\frac{\widetilde{\omega}_\alpha+\omega_\alpha}{2\sqrt{\widetilde{\omega}_\alpha \, \omega_\alpha}} \, b_\alpha+\frac{\widetilde{\omega}_\alpha-\omega_\alpha}{2\sqrt{\widetilde{\omega}_\alpha \, \omega_\alpha}}\, b_\alpha^\dagger
\end{equation}
where $\widetilde{\omega}_\alpha=\sqrt{\omega_\alpha^2+\omega_{P\alpha}^2}$ is the plasma-shifted transition frequency, with  $\omega_{P\alpha}^2=\dfrac{\strut 2 e^2 \, \Delta N_\alpha \, \omega_{\alpha} }{\strut \hbar \varepsilon_0\varepsilon_s}S_{\alpha \alpha}$ the plasma frequency.~\cite{nota_plasma} Note that, as already noticed by Helm,~\cite{helm} the plasma frequency does not depend on the dipole matrix element of the transition.

The term $H_{\rm{plasmon}}$ can then be written in terms of intersubband plasmons and their coupling as:
\begin{equation}
\label{plasma_hamiltonian}
H_{\rm{plasmon}}=\sum_\alpha{\hbar \widetilde{\omega}_\alpha p_\alpha^\dagger \, p_\alpha}+\frac{\hbar}{2} \sum_{\alpha \neq \beta}{\Xi_{\alpha,\beta} \Big(p_\alpha+p_\alpha^\dagger \Big) \Big(p_\beta+p_\beta^\dagger \Big)}
\end{equation}
The second term of $H_{\rm{plasmon}}$ expresses the dipole-dipole interaction between intersubband plasmons associated with different transitions,~\cite{delteil_APL} characterized by a coupling strength:
\begin{equation} \label{eq:Xi}
\begin{array}{l}
\Xi_{\alpha\beta} = \dfrac{\strut \omega_{P\alpha}\omega_{P\beta}}{\strut 2\sqrt{\widetilde{\omega}_\alpha\widetilde{\omega}_\beta}}\,C_{\alpha\beta}\\
C_{\alpha\beta} = \dfrac{S_{\alpha\beta}}{\sqrt{S_{\alpha\alpha}S_{\beta\beta}}}
\end{array}
\end{equation}
Expression~\ref{eq:Xi} shows that the coupling strength depends not only on the characteristics of the individual intersubband plasmons, but also on the overlap between the corresponding microscopic currents $C_{\alpha\beta} $, which is comprised between -1 and 1.

The coupling between $N$ intersubband plasmons produces $N$ new collective excitations, with eigenfrequencies $W_n$, which can be computed by numerically diagonalizing the following $2N \times 2N$ matrix~\cite{todorov_PRB1, todorov_PRB2}:
\begin{equation} \label{eq:matrixMgeneric}
 \mathbf{M} = \begin{pmatrix}
\widetilde{\omega}_1 & 0 & \Xi_{12} & -\Xi_{12} & &\Xi_{1N} & -\Xi_{1N}\\
0 & -\widetilde{\omega}_1 & \Xi_{12} & -\Xi_{12} &\cdots &  \Xi_{1N} & -\Xi_{1N}\\
\Xi_{12} & -\Xi_{12} & \widetilde{\omega}_2 & 0& &  \Xi_{2N} & -\Xi_{2N} \\
\Xi_{12} & -\Xi_{12} & 0 & -\widetilde{\omega}_2 & &  \Xi_{2N} & -\Xi_{2N}\\
 & &\vdots & & \ddots & & \\
\Xi_{1N} & -\Xi_{1N} &  \Xi_{2N} & -\Xi_{2N} & & \widetilde{\omega}_N & 0\\
\Xi_{1N} & -\Xi_{1N} &  \Xi_{2N} & -\Xi_{2N}&\cdots &0 & -\widetilde{\omega}_N\\
\end{pmatrix}
\end{equation}

Each new eigenmode of the system $W_n$ is associated with the excitation of a multisubband plasmon, described by the operators:
\begin{equation}
\label{eq:P_MSP}
P_n=\sum_\alpha{\left(a_{n\alpha}\, p_\alpha+b_{n\alpha} \, p_\alpha^\dagger\right)}
\end{equation}
where $a_{n\alpha}$ and $b_{n\alpha}$ are the components of the eigenvectors $\mathbf{V}_{n}$ of the matrix $\mathbf{M}$, written in the form:
\begin{equation}
 \mathbf{V}_{n} = \begin{pmatrix}
a_{n1}\\
b_{n1}\\
\vdots\\
a_{nN}\\
b_{nN}\\
\end{pmatrix}
_{2N\times1}
\end{equation}
with the normalization condition $\sum_i{\left( |a_{ni}|^2-|b_{ni}|^2\right)}=1$ which preserves the bosonic character of the collective excitations.

At this stage the Hamiltonian~(\ref{plasma_hamiltonian}) is written in terms of independent multisubband plasmon modes as:
\begin{equation}
H_{\rm {plasmon}}=\sum_n {\hbar W_n P_n^\dagger P_n}
\end{equation} 

\subsection{Multisubband plasmon current density and absorption spectrum}
In order to calculate the absorption spectrum in the presence of dipole-dipole interaction among intersubband plasmons, we use the same approach as in section~\ref{sec:ISB_transition}. We write the intersubband polarization in terms of the multisubband plasmons. The corresponding current densities are then calculated and related to the absorption spectrum of the electron gas in the presence of dipole-dipole interaction.  

Let us define the $N \times N$ matrix inverse
\begin{equation}
 X_{\alpha n} = \left( a_{n \alpha}+b_{n \alpha}\right)^{-1}
\end{equation}
By using Eq.~(\ref{eq:P_MSP}), the intersubband polarization operator reads:
\begin{equation}
\widehat{P}_z=\frac{e\hbar}{2m^*\sqrt{S}}\sum_n {\sum_\alpha {\frac{\xi_\alpha(z) \sqrt{\Delta N_\alpha}}{\sqrt{\omega_\alpha \widetilde{\omega}_\alpha}}} \,  X_{\alpha n} \Big( P_n+P_n^\dagger \Big)}
\end{equation}

The multisubband current density can now be calculated as the time evolution of the polarization under the Hamiltonian $H_{\rm{plasmon}}$:
\begin{equation}
\widehat{J}_z=\frac{i}{\hbar} \left[\widehat{P}_z,H_{\rm{plasmon}} \right]=i\sum_n{J_n(z)\Big(P_n-P_n^\dagger \Big)}
\end{equation}
with the spatial distribution of the multisubband plasmon given by
\begin{equation} \label{eq:def_MSP_J}
J_n(z)=\frac{e\hbar}{2m^*\sqrt{S}}W_n \sum_\alpha {\frac{\xi_\alpha(z) \sqrt{\Delta N_\alpha}}{\sqrt{\omega_\alpha \widetilde{\omega}_\alpha}}\,X_{\alpha n}}
\end{equation}

The absorption coefficient is obtained by integrating the contributions of the different multisubband plasmon current densities. Indeed, the absorption is related to the polarization of the medium, induced by the different intersubband transitions and in the presence of the depolarization field. The absorption coefficient is hence written as: 
\begin{eqnarray}\label{eq:abs_MSP_J}
\alpha_{2D}(\omega)&=&\frac{S}{\epsilon_0 c \sqrt{\epsilon_s}} A(\omega) \\ \label{eq:abs_MSP_J_2}
A(\omega)&=& \sum_n{\frac{1}{W_n} \left \vert \int_{-\infty}^{+\infty}{J_n(z)\mathrm{d}z} \right \vert^2  \textit{L}(\omega-W_n)}
\end{eqnarray}
as also shown in Appendix.\\
This quantity can also be expressed in terms of the eigenvalues and eigenvectors of the matrix $\mathbf{M}$, together with the characteristics of the individual intersubband transitions:
\begin{align}\nonumber
A(\omega) &=\sum_n  W_n \left \vert \sum_\alpha \sqrt{\Delta N_\alpha} \sqrt{\frac{\omega_\alpha}{\widetilde{\omega}_\alpha}} z_\alpha X_{\alpha n} \right \vert^2  \textit{L}(\omega-W_n)=\\ \label{eq:abs_prop_MSP}
&= \sum_n W_n F_n \textit{L}(\omega-W_n) 
\end{align}
$ W_n F_n$ can be considered as an effective oscillator strength of the multisubband plasmon modes (the multisubband equivalent of $f_\alpha \Delta N_\alpha$).
It is important to underline that each effective oscillator strength results from the contribution of all the optically active intersubband plasmons. They are weighted by the different quantities associated with individual transitions (dipole matrix element, transition frequencies) but also depend on the coupling between intersubband plasmons, which enters through the eigenvectors of the matrix $\mathbf{M}$ ($a_{J \alpha}$ and $b_{J \alpha}$).

The coupling between intersubband plasmons results in a redistribution of the absorption amplitude from the intersubband transitions to the multisubband plasmon modes. The total absorption satisfies the following sum rule:
\begin{equation}
\sum_\alpha {\omega_\alpha \left \vert z_\alpha \right \vert^2 \Delta N_\alpha}=\sum_n {W_n F_n}.
\end{equation}
This relation expresses the conservation of the total transition probability. It stems from the fact that the dipole - dipole interaction term $H_p$ commutes with the light - matter interaction term $H_{l-p}$.

\section{Absorption spectrum of a quantum well with several occupied subbands}\label{sec:single_well}
As a first application of our model, we consider a single quantum well with several occupied subbands. This system has been experimentally studied in Ref.~\onlinecite{delteil_PRL2012}, where it has been demonstrated that even though several intersubband transitions are optically excited, the absorption spectrum displays a single peak, concentrating the whole interaction with light. We will show that our model recovers this result.

\begin{figure}[htbp]
\includegraphics[width=\columnwidth]{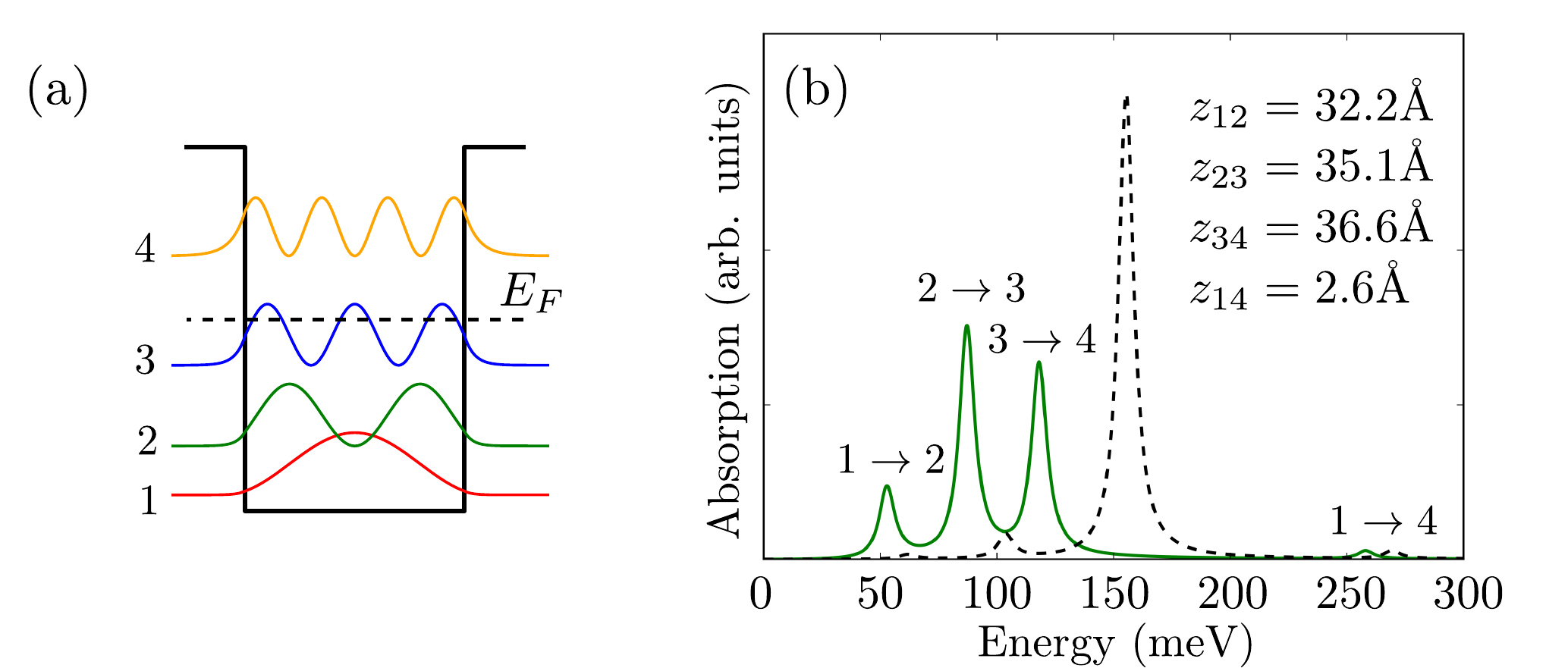}
\caption{a) Calculated band structure, energy levels and square moduli of the wave functions of a GaAs/Al$_{0.45}$Ga$_{0.55}$As quantum well, 15 nm wide, doped $N_v = 7 \cdot 10^{18}$ cm$^{-3}$. The corresponding Fermi energy at 0K is indicated with a black dashed line. b) Green line: Calculated absorption spectrum in single particle picture. Values of optical dipoles $z_{ij}$ of the transitions corresponding to the various peaks are shown. Black line: Absorption spectrum calculated with our model. The phenomenological broadening of the transitions is 8~meV.}
\label{fig:single_QW_ISB}
\end{figure}

Let us consider a 15 nm GaAs/Al$_{0.45}$Ga$_{0.55}$As quantum well, n-doped with a density $7 \times 10^{18}$~cm$^{-3}$, such that three subbands are occupied (see Fig.~\ref{fig:single_QW_ISB}(a), where the Fermi level is indicated by a dashed line). Figure~\ref{fig:single_QW_ISB}(b) presents the single particle absorption spectrum (green line), calculated as $A(\omega)=\sum_\alpha {\omega_\alpha \left \vert z_\alpha \right \vert^2 \Delta N_\alpha \textit{L}(\omega-\omega_\alpha)}$. The values of the dipole matrix elements of the four optically active transitions are also shown.

\begin{figure}[htbp]
\includegraphics[width=\columnwidth]{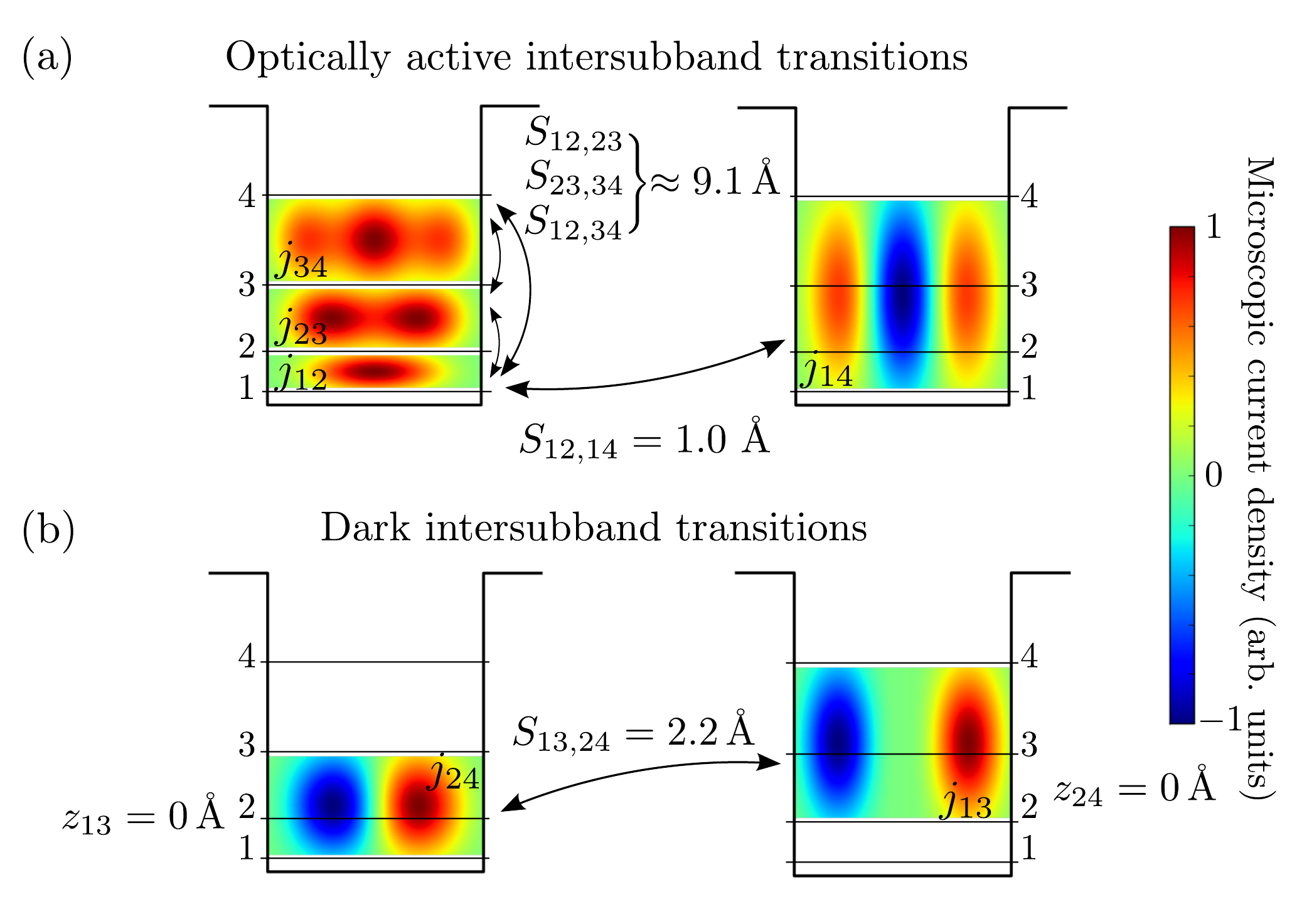}
\caption{Calculated spatial distributions of the microscopic current densities $j_{ij}$ for optically active (a) and dark (b) transitions in the 15 nm GaAs/Al$_{0.45}$Ga$_{0.55}$As quantum well of Fig.~\ref{fig:single_QW_ISB}(a). The distributions are plotted between the level energies they refer to and multiplied by a Gaussian function for visualization purposes. Coulomb lengths $S_{ij, kl}$ are also shown.}
\label{fig:microcurrents}
\end{figure}

Figure~\ref{fig:microcurrents} presents the spatial extension of the calculated microscopic current densities. Each microcurrent, normalized to its maximum amplitude, is reported in color scale as a function of $z$ and plotted in the energy interval between the confined levels involved in the corresponding transition. The microcurrents associated with dipole-allowed transitions are symmetric with respect to the center of the quantum well, as shown in panel (a), while microcurrents whose dipole is zero for parity are odd [panel (b)]. Note that transition 1-4 is optically active, even though $z_{14}$ is less than 10\% of $z_{i,i+1}$, $i=1,2,3$.

Figure~\ref{fig:microcurrents} also indicates the non-zero Coulomb lengths $S_{\alpha\beta}$ for some pairs of transitions. The highest values, $\approx 9$~\AA, are obtained for pairs of transitions between consecutive subbands, regardless of the quantum number of the subbands involved [Fig.~\ref{fig:microcurrents}(a)]. All these values are exactly equal in the case of an infinite quantum well, where $S_{n,n+1; m,m+1}=L_{\rm QW} /(2\pi^2)$ for all $n,m$. \\
Smaller values of $S_{\alpha,\beta}$ are instead obtained when the transition 1-4 is involved, indicating the existence of different scales of Coulomb lengths in the same quantum well. As an example, in Fig.~\ref{fig:microcurrents} we report the value of $S_{12,14}$, the highest one. Note that $S_{\alpha\beta}$ is also non negligible when the microcurrents involved present the same parity, even when they correspond to optically forbidden intersubband transitions, like 1-3 and 2-4 [Fig.~\ref{fig:microcurrents}(b)].

Table~\ref{tabella_C} presents the calculated values of the normalized overlap  between intersubband currents $C_{\alpha, \beta}$ for all the possible pairs of transitions. $C_{\alpha, \beta}$ is very close to 1 whenever transitions between consecutive subbands are involved, and it is in general non negligible for microcurrents presenting the same parity. This reflects the behavior already mentioned for $S_{\alpha,\beta}$.

\begin{table}[b]
\centering
\begin{tabular}{c|ccc|ccc}
(i,j) & (1,2) & (2,3) & (3,4) & (1,3) & (1,4) & (2,4)\\
\hline
(1,2) & 1 &0.93 & 0.93 &  0 & -0.28 & 0\\
(2,3) & 0.93 & 1 & 0.97 & 0 & 0.09 & 0\\
(3,4) & 0.93 & 0.97 & 1 & 0 & 0 & 0 \\
\hline
(1,3) & 0 & 0 & 0 &1  &0 & 0.84\\
(1,4) & -0.28 & 0.09 & 0 & 0 & 1 & 0 \\
(2,4) & 0 &0 &0 & 0.84  &0 &1\\
\hline
\end{tabular}
\caption{Calculated values of the normalized overlap $C_{\alpha,\beta}$ between microscopic currents.}
\label{tabella_C}
\end{table}

\begin{figure}[htbp]
\includegraphics[width=\columnwidth]{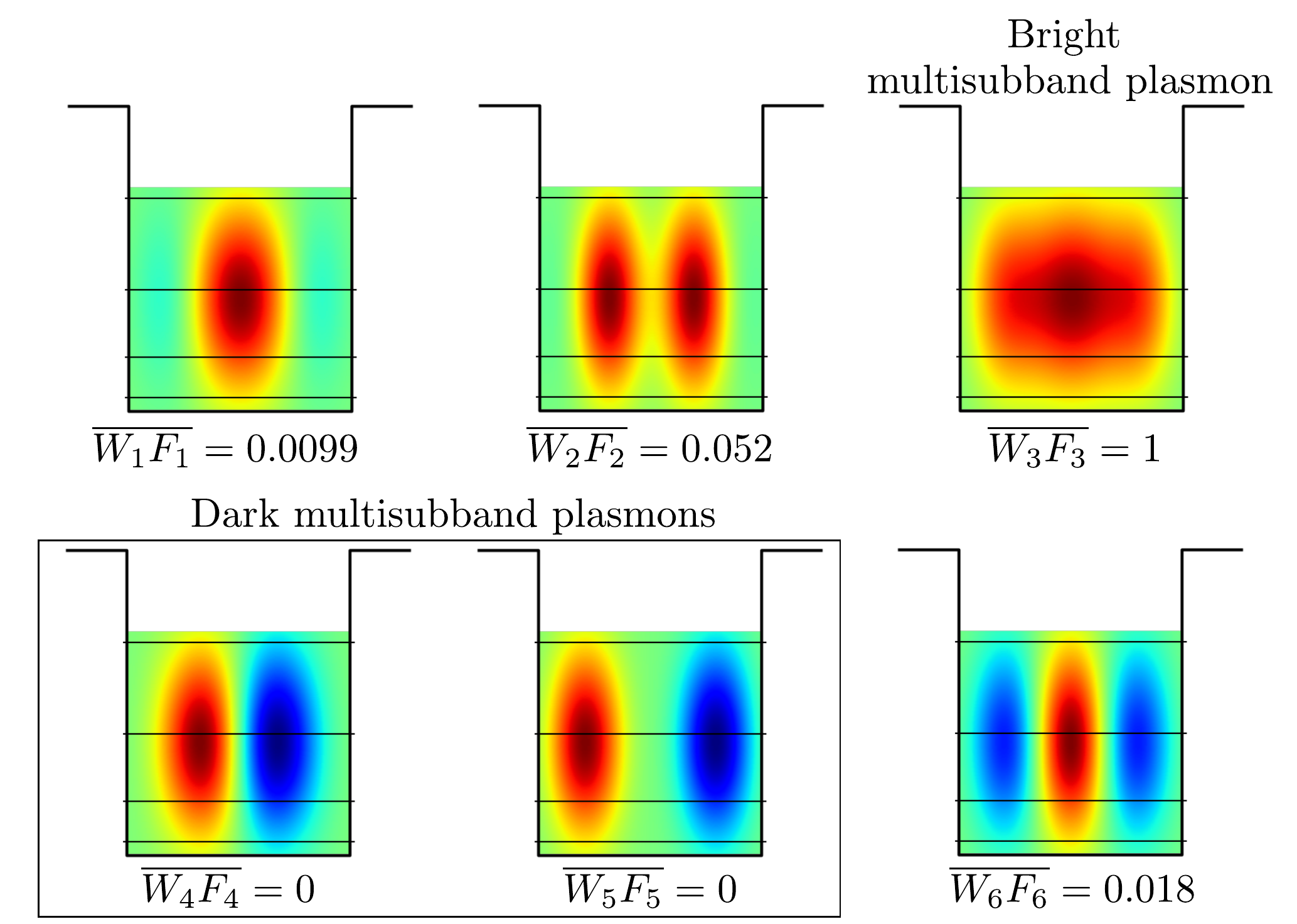}
\caption{Calculated spatial distributions of the microscopic collective current densities $J_{n}$ [Eq.~(\ref{eq:def_MSP_J})] in the 15 nm GaAs/Al$_{0.45}$Ga$_{0.55}$As quantum well of Fig.~\ref{fig:single_QW_ISB}(a). The distributions are plotted in the whole well energy range, and multiplied by a Gaussian function for visualization purposes. Values of effective oscillator strengths $\overline{W_nF_n} =\frac{W_nF_n}{\max\{W_nF_n\}}$ are also shown, with $W_nF_n$ as defined by Eq.~(\ref{eq:abs_prop_MSP}).}
\label{fig:J_MSP}
\end{figure}

The mutual coupling between the six intersubband transitions in this system leads to six multisubband plasmons. The corresponding current densities $J_n(z)$ are plotted in Fig.~\ref{fig:J_MSP}, expressing the spatial extension of the macroscopic polarization associated with each collective mode. In addition in fig.~\ref{fig:J_MSP} we indicate the effective oscillator strengths, normalized to the maximum value ($\overline{W_nF_n}$). It is evident from Fig.~\ref{fig:J_MSP} that two of the multisubband plasmons are dark, as the corresponding current densities are odd with respect to the center of the quantum well. Furthermore, even though there are four modes with a non-zero effective oscillator strength, we can consider that just one is {\em{bright}}, as its effective oscillator strength, $W_B F_B$, is at least 15 times greater than the others. This means that a macroscopic polarization is established in the quantum well only for this state, issued from the superposition in phase of all the different intersubband oscillators. This unique bright mode has the properties of a superradiant state~\cite{gross}. It is possible to estimate an effective dipole of the bright multisubband plasmon, $Z_B$. For this, we first observe that, due to Pauli blocking, in a quantum well with several occupied subbands not all the electrons in the system are involved in the interaction with light, but only a fraction corresponding to the occupation of the first subband: $N_{eff}=N_1$. As a consequence $W_B F_B= N_1 W_B \left \vert Z_B \right \vert ^2$. In the present example we obtain $Z_B=25.3$~\AA, corresponding to an oscillator strength of 1.75.  

\begin{figure} [htbp]
\includegraphics[width=\columnwidth]{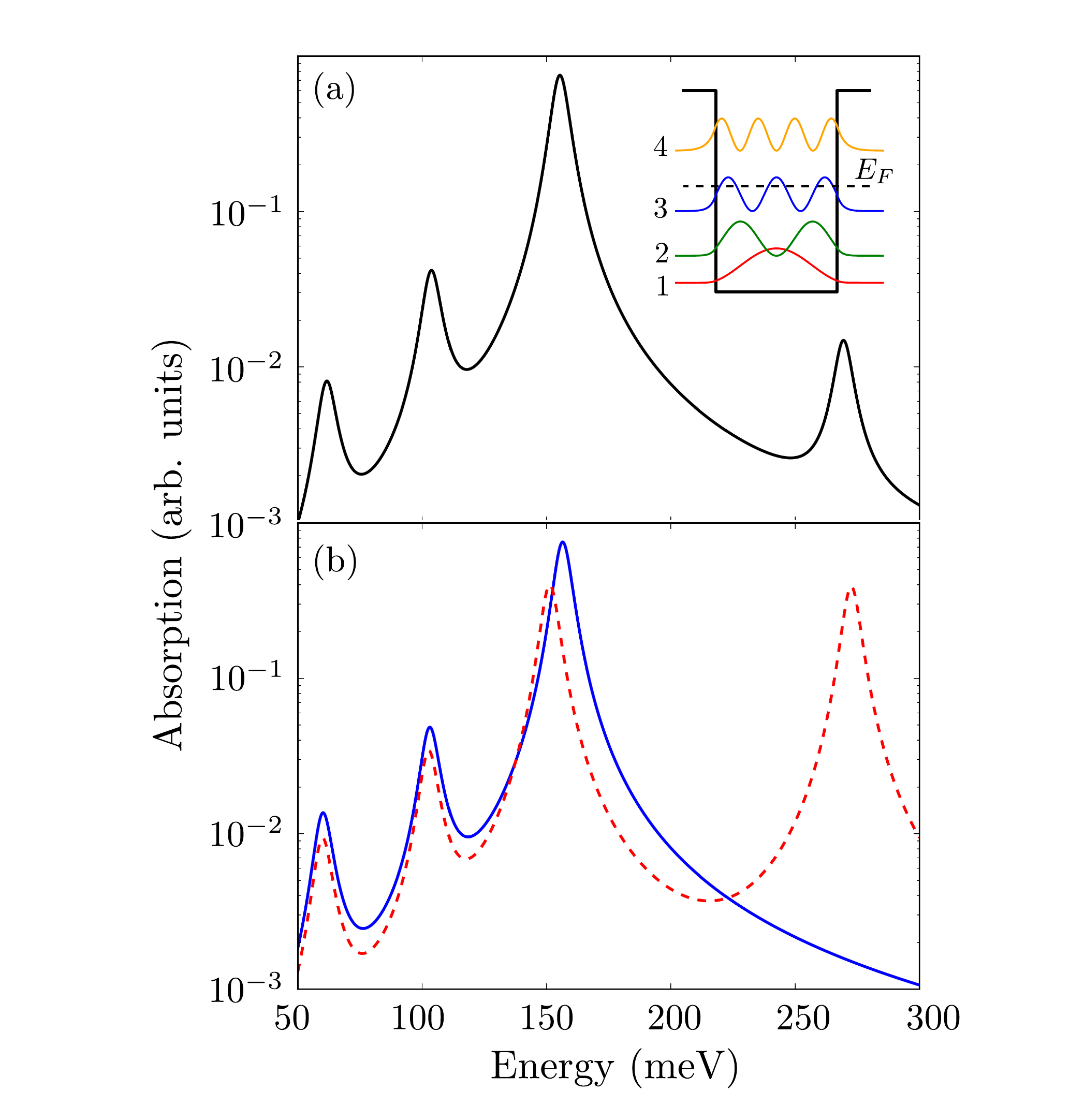}
\caption{a) Inset: Calculated band structure, energy levels and square moduli of the wave functions of a GaAs/Al$_{0.45}$Ga$_{0.55}$As quantum well, 15 nm wide, doped $N_v = 7 \cdot 10^{18}$ cm$^{-3}$. The corresponding Fermi energy at 0K is indicated with a black dashed line.  Main panel: Absorption spectrum in logarithmic scale calculated by using our quantum model. b) Blue line: absorption spectrum calculated by using Eqs.~(\ref{eq:ando})-(\ref{eq:abs}). Red line: absorption spectrum calculated by extending the sum in Eq.~(\ref{eq:ando}) to all optically active transitions.}
\label{fig:abs150}
 \end{figure}

The absorption spectrum  calculated with our model is shown in linear scale in Fig.~\ref{fig:single_QW_ISB}(b) (black line) and in logarithmic scale in Fig.~\ref{fig:abs150}(a). It displays four resonances at the energies of the multisubband plasmons, whose amplitude is proportional to their effective oscillator strength. The optical response of the electron gas is almost completely concentrated into a single peak, which has an amplitude more than one order of magnitude greater than the other resonances, in agreement with the experimental results obtained in Ref.~\onlinecite{delteil_PRL2012}.

Note that in the quantum well studied here the coupling with light is mainly determined by intersubband transitions between consecutive levels, mutually interacting on the same length scale, as determined by the value of $S_{\alpha,\beta}$ (see Fig.~\ref{fig:microcurrents}). The normalized overlap between the corresponding microcurrents is very close to one, as it can be seen in Table~\ref{tabella_C}. In this case, it is possible to calculate a dielectric function by using the following expression:~\cite{delteil_PRL2012, warburton_PRL}
\begin{equation}
\varepsilon(\omega)=\varepsilon_s \left( 1- \sum_{n=1}^{n_{\rm occ}} {\frac{\omega_{P_{n,n+1}}^2}{\omega^2-\omega_{n,n+1}^2+i\gamma \omega}}\right)
\label{eq:ando}
\end{equation}
where $n_{\rm occ}$ is the number of occupied subbands. The absorption spectrum (in cm$^{-1}$) accounting for the depolarization field is then calculated as:~\cite{helm,ando}
\begin{equation}
\label{eq:abs}
\alpha(\omega) = -\omega \frac{\sqrt{\epsilon_s}}{c} \mathrm{Im}\left[\frac{1}{\varepsilon(\omega)}\right]
\end{equation}

Equation~\ref{eq:ando} is a direct generalization of the dielectric function describing the excitation of a single intersubband transition: $\varepsilon(\omega) = \varepsilon_s \left(1-\frac{\omega_{P_{12}}^2}{\omega^2-{\omega}_{12}^2+ i\gamma\omega}\right)$. In this expression, each intersubband transition contributes to the permittivity through a Drude-Lorentz term, with a weight given by the corresponding squared plasma frequency. The absorption spectrum corresponding to Eq.~(\ref{eq:ando}) is presented as a blue line in Fig.~\ref{fig:abs150}(b). It is very similar to that calculated with our model and shown in panel (a) (except for the absence of the peak at $\approx 270$~meV, corresponding to energy $\hbar \widetilde{\omega}_{14}$).

Equation~\ref{eq:ando} is quite useful, as it can be easily generalized to the case of non-parabolic subbands.~\cite{askenazi} Nevertheless, it is important to underline that it does not provide a general expression of the dielectric function: it can only be used when all the Coulomb lengths involved are equal. In order to clarify this point, the red line in Fig.~\ref{fig:abs150}(b) presents the absorption spectrum obtained by extending the sum in Eq.~(\ref{eq:ando}) to all the optically active transitions, in this case all those between consecutive subbands plus the 1-4 transition. In this curve, the amplitude of the peak at 270~meV is almost equal to that of the peak at 155 meV. The different amplitude assigned to the 270 meV peaks in the two models is due to the fact that in Eq.~(\ref{eq:ando}) all the dipoles interact with the same Coulomb length, while our model does take into account the different Coulomb lengths between dipoles (see Table~\ref{tabella_C}).

As a consequence, our model can be used to calculate the dielectric response of electron gases confined in arbitrary one-dimensional potentials, unlike Eq.~(\ref{eq:ando}).

\section{Absorption spectrum of a system of two tunnel-coupled asymmetric quantum wells}
\label{sec:at}
\begin{figure}[htbp]
\includegraphics[width=\columnwidth]{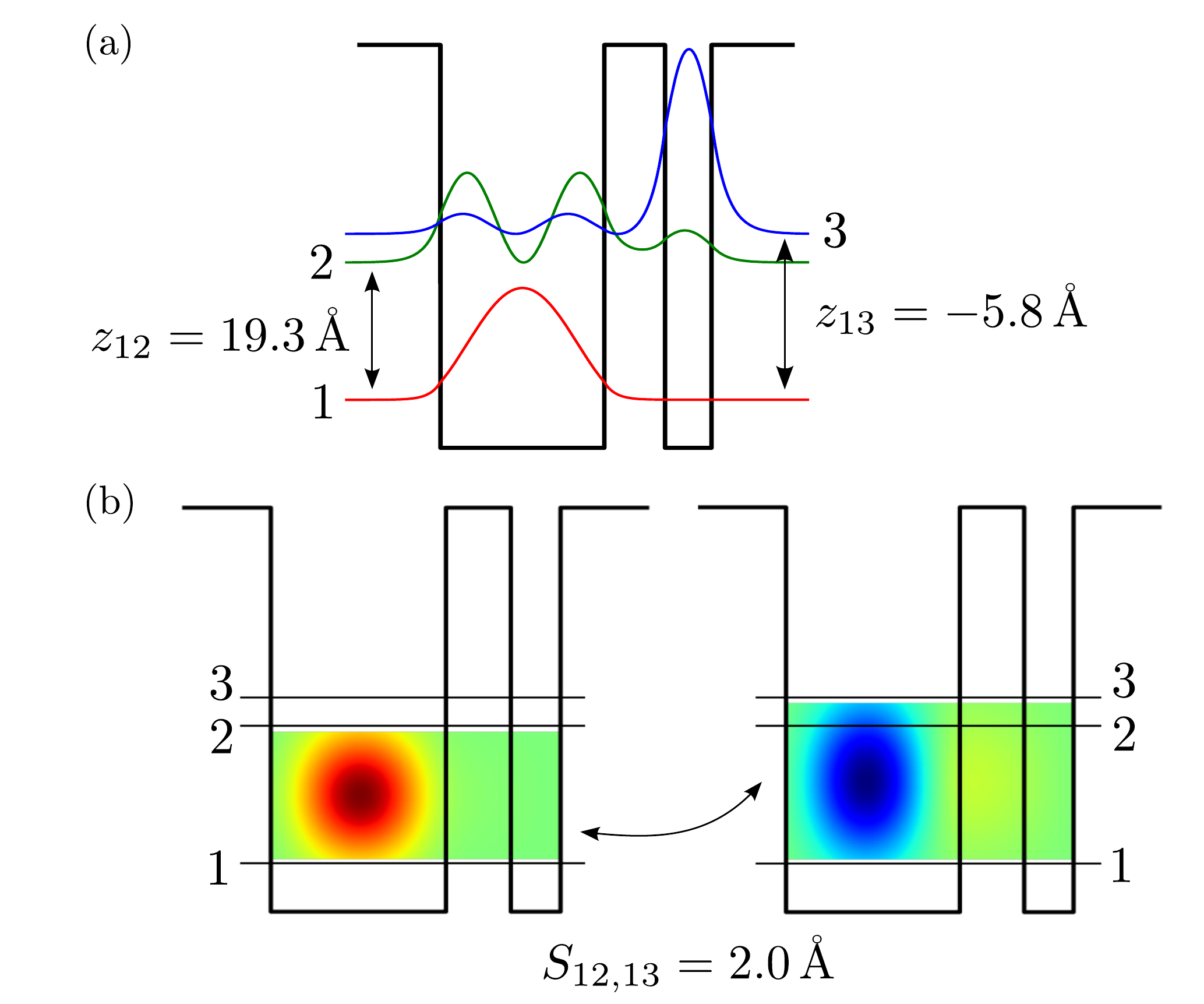}
\caption{a) Calculated band structure for coupled GaAs/Al$_{0.45}$Ga$_{0.55}$As quantum wells, with dimensions (in {nm}) 8.1/\textbf{3.0}/2.3. b) Calculated microscopic current densities $\xi_{i,j}$ for the two possible transitions 1-2 and 1-3. The distributions are plotted between the level energies they refer to and multiplied by a Gaussian function for visualization purposes.}
\label{fig:microcurrents_2}
\end{figure}

We consider as a second application of our model the calculation of the absorption spectrum for a system of two tunnel-coupled GaAs/Al$_{0.45}$Ga$_{0.55}$As quantum wells (8.1/3/2.3 nm). Figure~\ref{fig:microcurrents_2}(a) presents the band diagram and the square moduli of the wavefunctions in this system. The values of the dipole matrix elements are also reported: as the 1-3 transition is diagonal in the real space, its dipole is 70\% lower than the one of the 1-2 vertical transition.
Panel (b) presents the microscopic current densities (normalized to their maximum amplitude) of the 1-2 and 1-3 transitions. They are mostly localized in the largest quantum well. As a consequence, their Coulomb length is non negligible, resulting in an overlap factor close to one in modulus ($C_{12,13}=-0.95$).

\begin{figure}[htbp]
\includegraphics[width=\columnwidth]{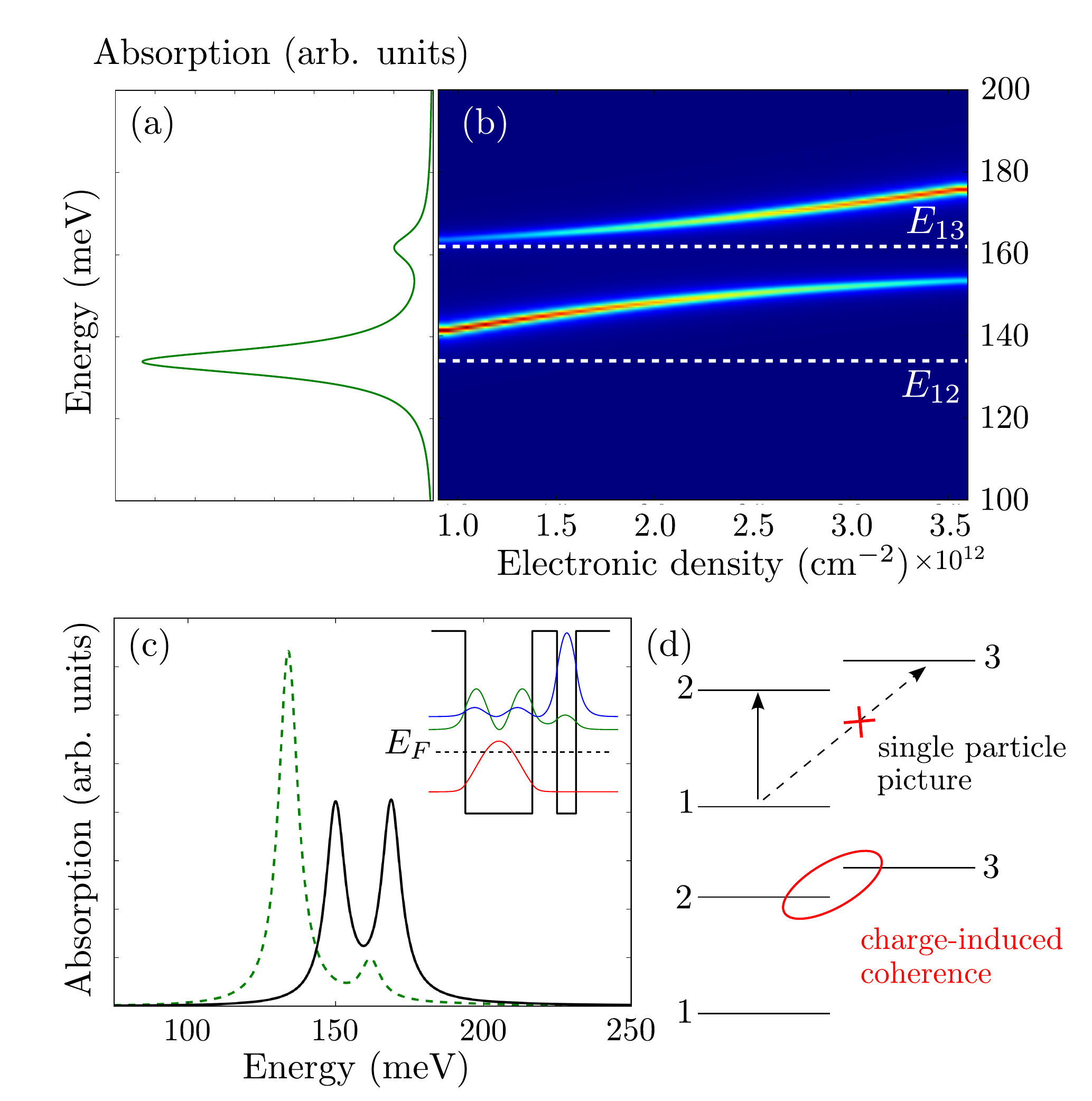}
 \caption{a) Calculated absorption spectrum for the coupled well structure presented in Fig.~\ref{fig:microcurrents_2}(a) in the single particle picture. b) Calculated absorption spectrum with our model for different values of electronic densities on the first subband. c)  Absorption spectrum calculated by using our model (black line) with $N_s= 2.4 \cdot 10^{12}$ cm$^{-2}$ (black line), compared with the single particle absorption spectrum (green line). Inset: calculated band structure for coupled GaAs/Al$_{0.45}$Ga$_{0.55}$As quantum wells, with dimensions (in nm) 8.1/\textbf{3.0}/2.3. d) Level energies schematization of the charge-induced coherence between levels 2 and 3, reminiscent of the Autler-Townes effect.}
 \label{fig:absAT}
 \end{figure}

The single particle absorption spectrum corresponding to this structure is presented in Fig.~\ref{fig:absAT}(a). It shows that, in this picture, the interaction with light is almost completely concentrated in the 1-2 transition peak, due to the difference in the dipole matrix elements between the vertical and the diagonal transition. 

The absorption spectrum calculated by using our model is shown in Fig.~\ref{fig:absAT}(b), in color scale, as a function of the electronic density in the first subband. The single particle transition energies are indicated by dashed lines. It is apparent that with the increasing electronic density, dipole-dipole Coulomb interaction redistributes the absorption amplitude between two peaks, at different energies with respect to the single particle ones. In particular, for a density $\approx 2.4 \times 10^{12}$~cm$^{-2}$ the two peaks have the same amplitude. The absorption spectrum calculated for this electronic density is presented in Fig.~\ref{fig:absAT}(c) (black line) and compared to the single particle one (green line).

The activation of the 1-3 transition, almost dark in single-particle picture, can be seen as the manifestation of the microscopic oscillation with a common phase of all the intersubband dipoles. This phenomenon is reminiscent of the Autler-Townes effect,~\cite{at_pr100_1955, cohen_book} which is a general quantum-mechanical effect observed on three-level quantum systems (cascade, lambda, or vee configurations) presenting an allowed 1-2 and a forbidden 1-3 transition.~\cite{imamoglu} When an intense and coherent coupling field at a frequency close to that of the 2-3 transition is shone on the system, the absorption of a second probe field (weak) presents a splitting. The splitting observed in the absorption spectrum in Fig.~\ref{fig:absAT}(b) can thus be seen as the signature of a laser-free Autler-Townes effect, in which the external coupling field is replaced by the charge-induced coherence [see scheme in in Fig.~\ref{fig:absAT}(d)].

Note that in this case the absorption spectrum predicted by our model coincides with the one calculated by using the expression derived by Allen et al.~\cite{allen, ando, chun} by using time dependent perturbation theory for the case of one occupied subband and two possible final states in an inversion layer.

\section{Tunnel coupling between multisubband plasmons}
\label{sec:coupledMPS}
Finally, let us consider two tunnel-coupled GaAs quantum wells, each of thickness $L=15$~nm,  identical to that presented in section~\ref{sec:single_well}, separated by a Al$_{0.45}$Ga$_{0.55}$As barrier.
The quantum wells are uniformly doped with an electronic density $7 \times 10^{18}$~cm$^{-3}$.

Two different lengths determine the properties of the system: the wavefunction extension in the barrier and the length of dipole-dipole interaction between intersubband plasmons. Our model allows the calculation of the optical properties of the electron gas by including these two characteristic lengths. The tunnel coupling is taken into account by considering the single particle eigenstates of the coupled well system, while the charge-induced coherence is included by calculating the coupling between all the intersubband plasmons associated with the transitions between extended states.
\begin{figure}[b]
\includegraphics[width=\columnwidth]{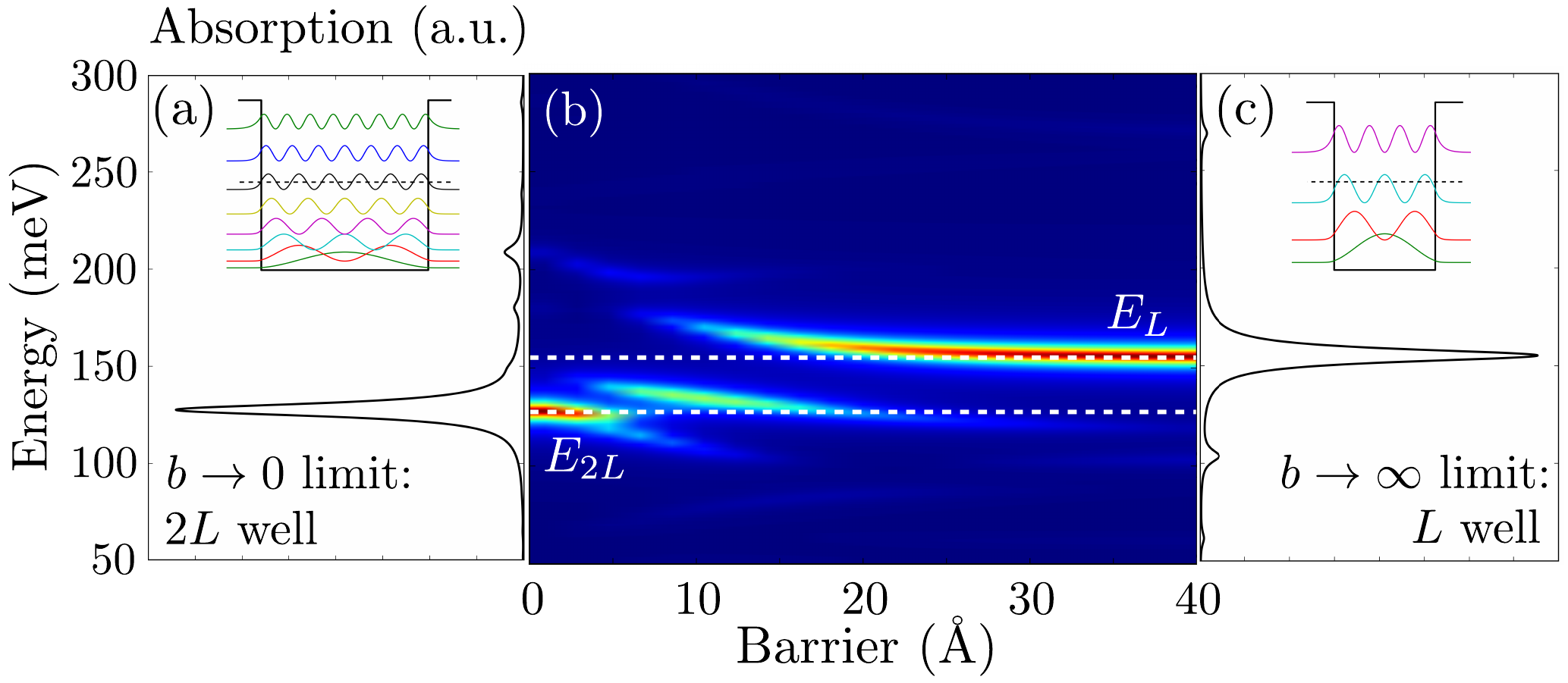}
\caption{Calculated absorption spectra for a 30 nm (panel (a)) and 15 nm (panel (c)) GaAs/Al$_{0.45}$Ga$_{0.55}$As quantum well. Inset: Band diagram and square moduli of the wavefunctions in the structure. b) Calculated absorption spectrum for a GaAs/Al$_{0.45}$Ga$_{0.55}$As structure 15/barrier/15 (in nm) for different values of the coupling barrier. }
\label{fig:coupled_plasmons1}
\end{figure}

The main panel of Fig.~\ref{fig:coupled_plasmons1} presents the calculated absorption  (in color scale) as a function of the energy and of the barrier thickness. In the limit of a large barrier, the absorption spectrum of a single well of thickness $L$ (presented in the right panel) is recovered. This is characterized by a main bright mode at energy $E_L$, indicated by a dashed white line. In the opposite limit, for a barrier thickness approaching zero, the spectrum shown in the left panel is obtained, which presents a single resonance at energy $E_{2L}$. \\
Note that $E_{2L}<E_{L}$: this can be understood by writing the energy of the multisubband plasmon in the single quantum well as:~\cite{askenazi}
\begin{equation}
E_{MSP}=\sqrt{E_P^2+E_{ISB}^2}
\end{equation}
Here $E_P^2=\sum_n{\left ( \hbar \omega_{P_{n,n+1}}\right)^2}$ is the effective plasma energy of the multisubband plasmon, and $E_{ISB}$ is the contribution of the confinement to the energy of the multisubband plasmon. The latter is given by the harmonic mean of the different intersubband transitions, weighted by the corresponding plasma energy: $E_{\rm ISB}=\hbar \sqrt{\frac{\Omega_P^2}{\sum_n{\frac{\omega_{P_n}^2}{\omega_n^2}}}}$. $E_{2L}$ is thus lower than $E_L$ because of the reduced transition energies in the wider well, as the plasma contribution $E_P = \hbar\Omega_P$ is the same in the two quantum wells.

The main panel of Fig.~\ref{fig:coupled_plasmons1} also shows that for barrier thicknesses between 5~{\AA} and 25~{\AA}, two absorption resonances with comparable amplitudes are observed. Interestingly, their energies present only a slight variation with the barrier thickness and stay close to $E_L$ and $E_{2L}$.

\begin{figure}[t]
\includegraphics[width=\columnwidth]{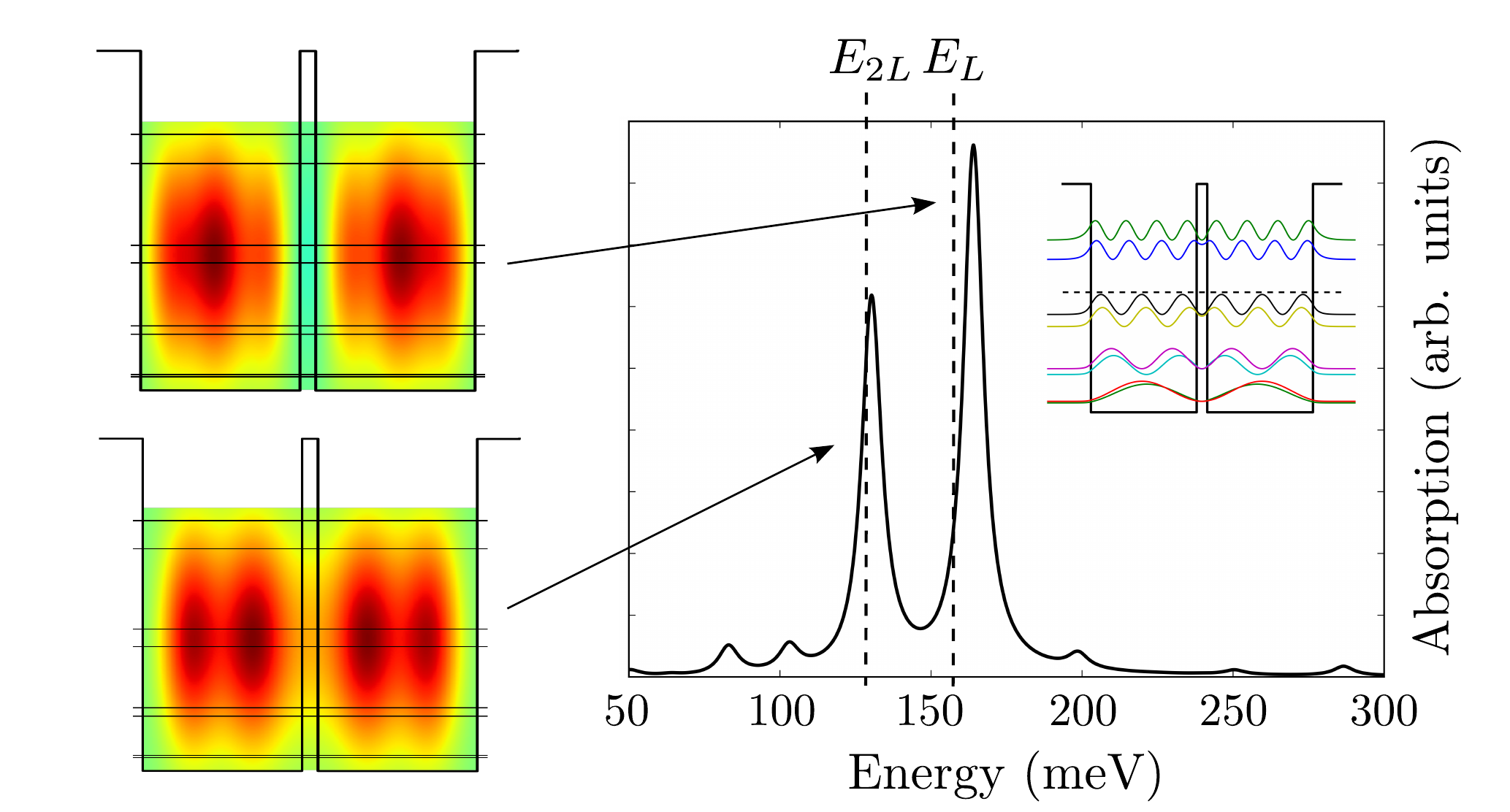}
\caption{On the right, calculated absorption spectrum for coupled GaAs/Al$_{0.45}$Ga$_{0.55}$As quantum wells of structure $L$/barrier/$L$ (in nm, 15/\textbf{1.5}/15). The two wells are uniformly doped, such that the electronic density is $N_v = 7 \cdot 10^{18}$ cm$^{-3}$. The vertical dashed lines indicate the resonances of quantum wells of width $L$ (15 nm) and $2L$ (30 nm). Inset: band diagram and square moduli of the wavefunctions in the structure. On the left, spatial distribution of microscopic collective currents corresponding to the two resonances (arbitrary units for the color scale). The distributions are plotted in the whole well energy range, and multiplied by a Gaussian function for visualization purposes.}
\label{fig:coupled_plasmons2}
\end{figure}

In order to understand the microscopic origin of these resonances and the role of tunnelling, let us fix the barrier thickness at 15~{\AA}. The calculated band diagram and the square moduli of the wavefunctions are presented in the inset of Fig.~\ref{fig:coupled_plasmons2}, showing the electronic doublets resulting from tunnel coupling. The absorption spectrum is presented in the main panel of Fig.~\ref{fig:coupled_plasmons2}, reporting two bright multisubband modes of comparable amplitude. Their closeness to the energies $E_L$ and $E_{2L}$ (indicated by dashed lines) suggests that tunnelling affects multisubband plasmons of adjacent quantum wells in an unconventional way. Instead of giving rise to a doublet around the uncoupled energy ($E_L$), it allows the coexistence of two single-well multisubband plasmons, one due to a $L=15$~nm well and the other to a $2L=30$~nm well. This observation is reinforced by the spatial distribution of the collective currents $J_n(z)$, also shown in Figure. The current associated with the peak at $E_L$ is composed of two lobes, localized in the individual quantum well. The multisubband current relative to the $E_{2L}$ peak still has two lobes, but in this case it is non zero also on the barrier.

The coexistence of the two absorption peaks is limited to a very short range of barrier thicknesses, between $\approx 10$ and 20 \AA. In fact, when the barrier is thin (below $\approx 5$ \AA) the wavefunctions of the quantum well of thickness $2L$ are only perturbatively modified. In the opposite limit, for barriers thicker than 20 \AA, the influence of tunnelling, still present on the shape of the wavefunctions, cannot be seen in the absorption spectrum. This is the result of two different contributions. Firstly, it is a manifestation of the oscillator strength transfer in favour of the high energy mode $E_{L}$, as already discussed in previous sections. Secondly, the coupling between microscopic current densities involves four wavefunctions (see the expression of $C_{\alpha,\beta}$ Eq.~\ref{eq:Xi}), while the tunnel coupling results from the overlap between two wavefunctions.    

\section{Conclusions}
In conclusion, we have presented a quantum formalism to calculate the absorption spectrum of a dense two-dimensional electron gas confined in an arbitrary potential, taking into account the effect of the depolarization field. The latter is included as a polarization self-interaction term in a  Hamiltonian written in the dipole representation. Our approach allows calculating the polarization of the electron gas in terms of microscopic current densities, whose spatial extension determines the optical properties of the system. The only requirement to perform the calculation is the knowledge of wavefunctions and electronic populations. We have presented three examples of the effect of the dipole-dipole coupling on the absorption spectrum of the system: a single quantum well displaying a unique collective bright state; a system of coupled asymmetric quantum wells, in which the charge induced coherence activates an almost dark intersubband transition; a system of symmetric highly doped quantum wells, showing an absorption doublet. A straightforward extension of our method allows the inclusion of the coupling with a cavity, and the study of ultra-strong coupling between multisubband plasmons and a microcavity mode.~\cite{delteil_PRL2012, todorov_PRB2, askenazi}
To the best of our knowledge, this work is the first theoretical investigation of the interplay between tunnel coupling and dipole-dipole interaction in the infrared optical properties of an electron gas. Our model could be used to conceive novel infrared quantum emitters, taking advantage of the superradiant character of the bright multisubband plasmon states.

\begin{acknowledgments}
We thank L. C. Andreani and F. Alpeggiani for fruitful discussions. This work has been supported by the ERC grant 'ADEQUATE'.
\end{acknowledgments}

\appendix*{}\label{appendix}
\section{Derivation of the absorption coefficient}
Let us consider the light-matter interaction term of the Hamiltonian~(\ref{int_hamiltonian}):
\begin{equation}
H_{l-p} =  - \frac{1}{\varepsilon_0 \varepsilon_s} \int P_zD_z \,{\rm d}^3 \mathbf{r} 
\end{equation}
where only the $z$ component is considered because of the intersubband absorption selection rule.
We adopt a fully quantum description, in which the electric displacement vector $D_z$ is expressed in terms of the creation $a_{\mathbf{q}}^\dagger$ and destruction $a_{\mathbf{q}}$ operators of a photon with wavevector $\mathbf{q}$ and frequency $\omega_{c,\mathbf{q}}$:
\begin{align} \label{eq:appendix2}
&P_z = \sum_{\alpha, \mathbf{q}} \frac{j_\alpha (z)}{\omega_\alpha} e^{i\mathbf{qr}} (b_{\mathbf{q} \alpha} +b^\dagger_{-\mathbf{q} \alpha})
\\
&D_z = i \sum_\mathbf{q} A_\mathbf{q} f_\mathbf{q}(z)\frac{|\mathbf{q}|}{\omega_{c\mathbf{q}}}e^{i\mathbf{qr}}(a_\mathbf{q} - a^\dagger_{-\mathbf{q}})
\\
&A_\mathbf{q} = \sqrt{\frac{\hbar \omega_{c\mathbf{q}}}{2 \mu_0 SL_\mathbf{q}}}
\\
&\int f^2_\mathbf{q}(z) \,{\rm d}z = L_\mathbf{q}
\end{align}
where  $f_{\mathbf{q}}(z)$ is an arbitrary guided mode with wavevector $\mathbf{q}$, $S$ is the area of the system and $L_{\mathbf{q}}$ the light-matter interaction length.\\
Then $H_{l-p}$ is evaluated to be:
\begin{align}\nonumber
H_{l-p} = &\sum_{\alpha, \mathbf{q}} C_\mathbf{q} \frac{i\sqrt{\omega_{c\mathbf{q}}}\sin \theta_\mathbf{q}}{\omega_\alpha} \int j_\alpha (z) \,{\rm d} z \,(b_{\mathbf{q} \alpha} +b^\dagger_{-\mathbf{q} \alpha})\\
&\times(a_\mathbf{q} - a^\dagger_{-\mathbf{q}})
\\
C_\mathbf{q} = &\sqrt{\frac{\hbar}{2\varepsilon_s \varepsilon_0 S L_\mathbf{q}}}f_\mathbf{q}(z_0) 
\end{align}
Here we applied the long wavelength approximation around the position $z_0$ of the quantum well system and $\sin \theta_\mathbf{q} = |\mathbf{q}|c/\sqrt{\varepsilon_s}\omega_{c\mathbf{q}} $ is the propagation angle inside the substrate.\\
Let us consider a perturbative regime where the external field is weak. Then we can consider an absorption process, where the electronic system is initially in its ground state, $|F \rangle$, and there is one photon. The initial state of the system is then $a^\dagger_\mathbf{q}|F \rangle$. The final state is no photon an intersubband excitation with an energy $\hbar \omega_\alpha =\hbar \omega_{c\mathbf{q}}$: $b^\dagger_{\mathbf{q} \alpha}|F \rangle$. Then according to Fermi's golden rule the absorption rate is proportional to:
\begin{eqnarray}
A_\alpha (\omega) =  |\langle F| a^\dagger_\mathbf{q} H_{l-p} b^\dagger_{\mathbf{q} \alpha}|F \rangle|^2 \delta (\omega - \omega_\alpha) \nonumber \\
= |C_\mathbf{q}|^2 \sin^2 \theta_\mathbf{q} \frac{1}{\omega_\alpha} \Big{|} \int j_\alpha(z)dz\Big{|}^2 \delta (\omega - \omega_\alpha)
\end{eqnarray}

Here $\omega = \omega_{c\mathbf{q}}$ is the frequency of the incident photon, and the factor $\sin^2 \theta_\mathbf{q}$ expresses the selection rule for polarization of the incoming radiation. Then Eq.~(\ref{eq:abs_spectrum_ISB}) in the main text is obtained by summing all possible excitations $\alpha$ and replacing the Dirac delta with broadened Lorentzians, and discarding the $\mathbf{q}$-dependence. Clearly, if we use the expression of the polarization field for multisubband plasmons:

\begin{equation}
P_z = \sum_{n, \mathbf{q}} \frac{J_n (z)}{W_n} e^{i\mathbf{qr}} (P_{\mathbf{q} n} +P^\dagger_{-\mathbf{q} n}) 
\end{equation} 

we will obtain Eq.~(\ref{eq:abs_MSP_J_2}), since the formal expression of the polarization field is the same as Eq.~(\ref{eq:appendix2}) above.

\newpage



\begin{thebibliography}{20}
\bibitem{nota_plasma} The plasma frequency is usually written in terms of the the effective length of interaction between the intersubband dipoles and the electromagnetic field: $L_{\rm eff}=\dfrac{\hbar f_{12}}{2 m^* \omega_{12} S}$, with $f_{12}$ the oscillator strength of the transition. In this case the plasma frequency reads: $\omega_{P_{12}}^2=\frac{e^2 N_s f_{12}}{\hbar \varepsilon_0 \varepsilon_s m^* L_{\rm eff}}$.
\end{thebibliography}
\end{document}